# Impacts of Nuclear War on Human Health from Changed Surface Ultraviolet Radiation

Shu Xu[1], Lili Xia[1], Alan Robock[1], Charles Bardeen[2], Simchan Yook[3], Sasha Madronich[4], and Soko Setoguchi[5,6]

[1] Department of Environmental Sciences, Rutgers University, New Brunswick, NJ, USA
[2] NSF National Center for Atmospheric Research, Boulder, CO, USA
[3] Department of Earth, Atmospheric and Planetary Sciences, Massachusetts Institute of Technology, Cambridge, MA, USA
[4] Madronich Scientific Consulting LLC, Boulder, CO, USA
[5] Rutgers Health-RWJBarnabas Health Center for Climate, Health, and Healthcare, New Brunswick, NJ, USA
[6] Department of Medicine, Robert Wood Johnson Medical School, Rutgers University, New Brunswick, NJ, USA

Corresponding author: Shu Xu (sx175@envsci.rutgers.edu)

Key Points:

- Nuclear war driven ozone loss could sharply increase surface ultraviolet radiation
- Enhanced ultraviolet radiation significantly shortens safe outdoor exposure times making unprotected activities highly dangerous
- Enhanced ultraviolet radiation causes massive amounts of extra skin cancer deaths and this severe impact will last for a century

Submitted to *Earth's Future*

## Abstract

Climate model simulations indicate that surface ultraviolet (UV) radiation would change after soot injection into the stratosphere during a nuclear war, due to the competing effects of ozone depletion and aerosol attenuation. Using climate model simulations, we evaluate UV impacts under two scenarios: a regional India-Pakistan conflict producing 5 Tg of soot and a global U.S.-Russia war producing 150 Tg. UV enhancements due to ozone depletion substantially shorten safe outdoor exposure time, particularly for individuals with lighter skin types. Applying UV dose-response relationships to the year 2000 population data, without accounting for direct conflict mortality or famine-related population loss, the 5 Tg scenarios result in approximately 5,300-9,800 additional skin cancer deaths within 10-15 years and up to 75,000 cumulative excess deaths over the following century. In contrast, strong aerosol attenuation under the 150 Tg scenario initially suppresses surface UV, resulting in about 8,500 fewer skin cancer deaths within 15 years and a maximum cumulative reduction of approximately 17,000 deaths over the following century. These findings demonstrate that nuclear war-induced changes in surface UV radiation represent a persistent but previously understudied health impact of nuclear war. While skin cancer would not dominate overall mortality following a nuclear war, enhanced cumulative surface UV radiation represents an additional, long-lasting threat to human health that compounds other global impacts such as climate disruption and food insecurity. The excess UV may also pose negative impacts on animals and plants, including those used for agriculture, which remain to be quantified.

## Plain Language Summary

A nuclear war would inject massive amounts of soot into the upper atmosphere, damaging the ozone layer and changing how much harmful ultraviolet (UV) radiation reaches Earth. This study simulates two scenarios, a regional war between India and Pakistan and a global war between the U.S. and Russia. We found that ozone damage allows more dangerous UV rays to reach the surface, drastically reducing the time people can safely stay outdoors without protection, especially for people with lighter skin. Using the year 2000 global population and not including deaths caused directly by the war or by famine, the increased UV exposure could cause about 10,000 additional skin cancer deaths within the first decade and more than 75,000 over the following century after a smaller regional war. But a global war would produce so much soot that it would initially block sunlight and reduce surface UV, leading to long-term reduction of skin cancer deaths even after soot falls out. While skin cancer would not be the main cause of total casualties after a nuclear conflict, this increased UV radiation represents a severe, long-lasting threat. It would compound other global crises like climate disruption and food shortages, while potentially harming the plants and animals we need for agriculture.

# 1 Introduction

Globally, the total inventory of nuclear weapons has declined since 1986 (Kristensen et al., 2025), but has leveled out for the past 15 years. The number of deployable warheads remains high—approximately 9,615 as of June 2025 (RECNA, 2025)—more than sufficient for a nuclear conflict to produce catastrophic consequences for Earth's climate and human society. A large-scale nuclear war between the United States and Russia could ignite widespread urban and industrial fires, producing vast quantities of smoke. These aerosols, particularly black carbon, would absorb solar radiation, heat the surrounding air, rise into the stratosphere, disperse globally, and persist for more than a decade (Robock, 2010). The resulting cool, dark, and dry surface conditions are commonly referred to as "nuclear winter" (Turco et al., 1983; Robock et al., 2007; Coupe et al., 2019).

Previous simulations support the hypothesis that surface ultraviolet (UV) radiation would increase following the injection of soot aerosols into the stratosphere. Mills et al. (2014) showed global ozone losses of 20-25%, and peak UV index (UVI) values at 12-21 in June over the most populous regions of North America and southern Europe after a regional nuclear war between India and Pakistan with 5 Tg black carbon injection. The UVI, adopted internationally, reflects the wavelength-dependent effectiveness of solar radiation in causing skin damage (McKinlay, 1987). According to the World Health Organization, sun protection is recommended when the UVI reaches 3 (e.g., seeking shade at midday, wearing protective clothing, using sunscreen and hats), while at a UVI of 8 additional precautions, such as avoiding outdoor exposure during midday, are advised. A UVI above 11 is classified as "extreme" (WHO, 2002).

Bardeen et al. (2021) used the same coupled climate model as Mills et al. (2014), with updated aerosol size assumptions and $NO_x$ emissions, to simulate 15 years following both a regional nuclear war and a global conflict involving 150 Tg of black carbon injection. In the regional scenario, they found a global column ozone reduction of about 25%, with recovery taking approximately 12 years. In the global U.S.-Russia war scenario, stratospheric aerosols initially reduce surface UV by blocking sunlight during the first several years. However, as soot heats the stratosphere and accelerates ozone depletion, and as black carbon is gradually removed and the shading effect weakens, peak global ozone losses reach ~75%, with reductions of ~65% in the tropics. This severe depletion allows substantially more UV radiation to reach the surface. Consequently, UVI values exceed 35 in the tropics 6-10 years after injection, and surpass 45 during summer in the southern polar regions 6-9 years after injection—far above the World Health Organization "extreme" threshold of 11.

Building on Bardeen et al. (2021), Yook et al. (2025) incorporated halogen emissions (chlorine and bromine) from urban fuels as well as a new heterogeneous chemistry scheme that accounts for hydrochloric acid solubility in organic carbon. Their simulations show that a regional war scenario with 5 Tg of soot could result in a ∼40% reduction in the global ozone burden, nearly doubling previous studies. Ozone losses exceeded ∼80% over the Arctic and reached ∼50% over Northern Hemisphere mid-latitudes, including highly populated areas, further amplifying surface UV increases.

Elevated surface UV radiation poses significant threats to human health. Solar UV radiation consists of UV-A (315-400 nm), UV-B (280-315 nm), and UV-C (100-280 nm). As sunlight passes through the atmosphere, all UV-C and approximately 90% of UV-B are absorbed by ozone layer, whereas UV-A is much less affected (WHO, 2002). As a result, ambient sunlight is dominated by UV-A (90-95%), with a smaller contribution from UV-B (5-10%), depending on latitude, season, and time of day. UV-A penetrates deeply into the dermis and generates reactive oxygen species that indirectly damage DNA, whereas UV-B is largely absorbed in the epidermis but can damage DNA by resulting molecular rearrangements and forming characteristic photoproducts that lead to mutations (D'Orazio et al., 2013). A given dose of UV-B is approximately 1000 times more effective at causing erythema (sunburn) than the same dose of UV-A. Consequently, despite its smaller proportion, UV-B is the primary driver of erythema (Young et al., 2011), which is a strong predictor of skin cancer (Gandini et al., 2005; Iannacone et al., 2012).

Skin cancers are the most common malignancies worldwide, with nearly 5.5 million cases diagnosed annually (Wehner, 2026; Joshi et al., 2025). In the United States alone, they account for over 20,000 deaths each year (Aggarwal, 2019) and impose medical treatment costs of $8.9 billion annually (Kao et al., 2023). Skin cancers are broadly classified into melanoma and non-melanoma skin cancers (NMSC) based on cellular origin and clinical behavior (D'Orazio et al., 2013). Extensive epidemiological and molecular evidence links all forms of skin cancer to UV exposure (Linos et al., 2009). It is estimated that UV radiation contributes to approximately 75% of melanoma cases and up to 83% of NMSC (Neale et al., 2023). However, relatively few studies have quantitatively linked UV-B dose to specific health outcomes such as sunburn and skin cancer (e.g., Slaper et al., 1996; van Dijk et al., 2013; Madronich et al., 2021), and the health impacts of enhanced UV radiation following a nuclear war remain poorly understood.

In this study, we used previous simulations of a regional India-Pakistan conflict (5 Tg soot) and a U.S.-Russia conflict (150 Tg soot) (Bardeen et al., 2021; Yook et al., 2025). Using year-2000 climate conditions as the control, the model was run to simulate 15 years by Bardeen et al. (2021) and 10 years by Yook et al. (2025) after soot injection. Our results indicate that sunburn risk would rise significantly during peak UV years, particularly for individuals with lighter skin types. We then estimate additional skin cancer mortality in the three largest nuclear-armed countries, the United States, China, and Russia. Even a regional India-Pakistan war would result in thousands of additional deaths. Moreover, excess UV exposure permanently increases an individual's lifetime skin cancer risk, leading to health impacts that persist for a century after the war and more than a fivefold increase in total mortality, even after UV levels return to baseline.

## 2 Materials and Methods

### *2.1 Data*

To assess the impacts of regional and global nuclear wars on the global climate and human society, Bardeen et al. (2021) used the Community Earth System Model (CESM) version 1 (Hurrell et al., 2013), a chemistry-climate model with interactive atmosphere, land, ocean, and sea ice components. The ocean model simulates the evolution of physical and biogeochemical parameters at 1° horizontal resolution with 60 vertical layers of varying depth by coupling the Parallel Ocean Program (POP) version 2 ocean physical model (Danabasoglu et al., 2012) with an ocean biogeochemical model (Lindsay et al., 2014). The land model is the Community Land Model (CLM) version 4 with a carbon-nitrogen cycle and simulates the evolution of the land physical state, characteristics of the land surface, exchanges of energy and material with the atmosphere, and run-off into the ocean. It has a horizontal resolution of 1.9° × 2.5° with 15 vertical layers for the land and 10 vertical layers for lakes (Bonan et al., 2011; Oleson et al., 2010). The atmospheric model used in their study is the Whole Atmosphere Community Climate model version 4 (WACCM4, Marsh et al., 2013), using a grid with 1.9° × 2.5° horizontal resolution, 66 vertical layers and a 140 km model top.

The Tropospheric Ultraviolet and Visible (TUV) model version 4.2 (Madronich and Flocke, 1997) was coupled to the system to simulate spectral UV radiation and biologically weighted UV doses based on their impacts on human health and ecosystems at each model grid cell. Yook et al. (2025) followed the framework of Bardeen et al. (2021) and improved the chemical representation by incorporating halogen and organic carbon emissions and updating heterogeneous chemistry processes.

In this study, we analyze existing simulations including a 15-year 5 Tg soot regional nuclear war scenario and a 15-year 150 Tg soot global nuclear war scenario from Bardeen et al. (2021), together with a 10-year 5 Tg soot regional nuclear war simulation with updated forcings and chemistry from Yook et al. (2025). The model outputs have a horizontal resolution of 1.9° × 2.5° and are analyzed at monthly temporal resolution.

### *2.2 Calculation of UV impact on sunburn damage*

The minimal erythema dose (MED) is widely used as a biological measure of UV exposure and reflects a person's sensitivity to UV radiation. MED is defined as the least amount of UV-B radiation that causes reddening and inflammation of the skin 24-48 hours after exposure (D'Orazio et al., 2013). In this study, we use MED to quantify the threshold of UV exposure at which an individual develops sunburn. The MED depends on the UV spectrum, skin type, body site, and season, and varies across UV sources, and here we adopt the estimates from D'Orazio et al. (2013), which quantified MED for people of different phenotype (further discussed in Section 2.3.3). The exposure time required to induce sunburn is therefore defined as:

$$t = \frac{MED}{UVB} \tag{1}$$

where MED is the minimum erythemal dose defined above, expressed in J $m^{-2}$, and UVB is the surface UV-B irradiance in W $m^{-2}$.

### *2.3 Calculation of UV impact on skin cancer*

#### *2.3.1 Biologically Weighted UV Radiation*

Solar UV radiation incident on Earth's surface is composed of wavelengths spanning the range of 280 nm to 400 nm. Over this same wavelength range, biological sensitivity to UV photons can vary by orders of magnitude (Madronich et al., 2021). Integrating over wavelengths λ, a biologically weighted irradiance $I_{bio}$ is defined as:

$$I_{bio} = \int_{280\ nm}^{400\ nm} I(\lambda)\ A_{bio}(\lambda)\ d\lambda \quad (2)$$

where $I(\lambda)$ is the sunlight's spectral irradiance and $A_{bio}(\lambda)$ is the action spectrum (or spectral sensitivity function) for skin cancer (Figure 1).

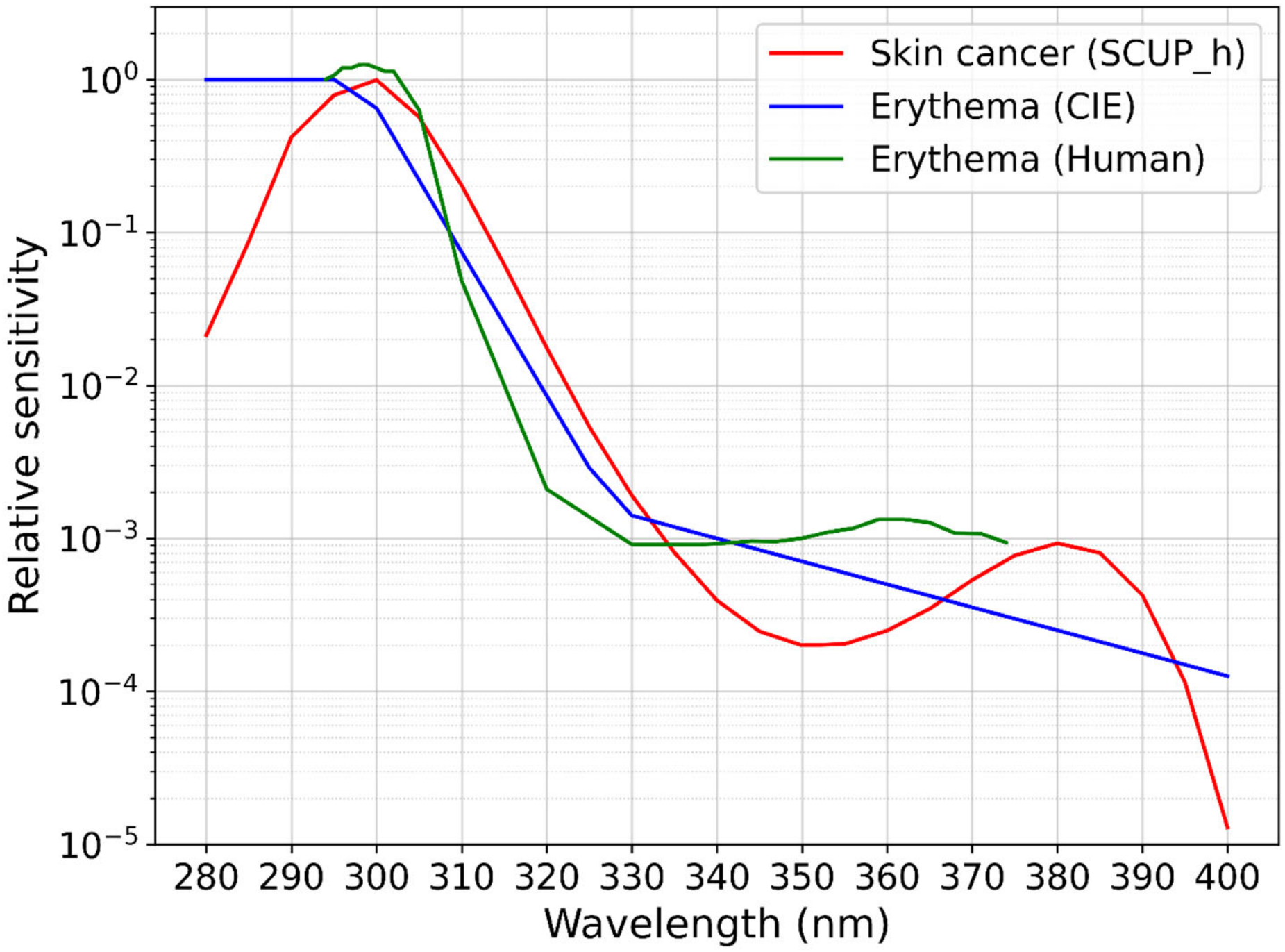

**Figure 1.** Spectral dependence (action spectrum) used in the TUV model for the induction of erythema (blue and green curves) and skin cancer (red curve). The skin cancer action spectrum applied in this study is derived from the Skin Cancer Utrecht/Philadelphia dataset, corrected for human transmission (SCUP-h), and represents the biological effectiveness for all major types of skin cancer (de Gruijl et al., 1993). The erythema action spectrum is also shown because it forms the basis for calculating the UV Index and assessing the skin-type-dependent response of the global population to UV radiation. Specifically, we include both the CIE (Commission Internationale de l'Eclairage or Technical Committee of the International Commission on Illumination) standard erythema weighting function (Webb et al., 2011) and the human skin-based erythema weighting function proposed by Anders et al. (1995).

*2.3.2 Dose-effect relations for skin cancer*

Excess skin cancer risk is quantified relative to changes in the annual biologically weighted UV dose at a given location. The excess risk represents the projected increase in skin cancer incidence associated with enhanced annual UV exposure, with the effects of even a few years of exposure persisting over a lifetime (Slaper et al., 2001). Changes in skin cancer incidence (number of new cases) or mortality (number of deaths), denoted as $\Delta I$, resulting from variations in UV radiation between time periods $t_1$ to $t_2$ are derived using the annual UV doses $UV\,(x, y, t_1)$ and $UV\,(x, y, t_2)$ and a reference UV dose ($UV_0$), as follows:

$$\Delta I = \sum_{i=1}^{3} I_i \left[ \left( \frac{UV(x, y, t_2)}{UV_0} \right)^{C_i} - \left( \frac{UV(x, y, t_1)}{UV_0} \right)^{C_i} \right] \quad (3)$$

where $I_i$ denotes baseline skin cancer incidence or mortality, and $i$ (as a subscript) indexes the three skin cancer types: basal cell carcinoma (BCC), squamous cell carcinoma (SCC), and melanoma. BCC and SCC together represent non-melanoma skin cancer (NMSC). The coefficients $C_i$ are derived from statistical regression analysis (Slaper et al., 1996).

In this study, following the methodology of Madronich et al. (2021), we define $t_1$ as the reference baseline period such that $UV\,(x, y, t_1) = UV_0$. Under this assumption, the relationship can be reformulated as a locally linear approximation. Accordingly, for each type of skin cancer, we obtain:

$$\frac{\Delta I}{I} = C \frac{\Delta UV}{UV_0} \quad (4)$$

where $\Delta UV$ represents the change in UV dose relative to the reference baseline period. The coefficient $C$ denotes the percentage change in skin cancer incidence or mortality associated with a 1% increase in UV dose. This coefficient is commonly referred to as the Biological Amplification Factor (BAF) (Madronich et al., 2021).

To project the responses of populations with different age groups and skin phototypes, the dose-effect relationship is further expressed as:

$$\Delta HEC(P_n) = \frac{BAF}{f_p} \times \sum_{(age=0:85)} \left[ \Delta UV_{exp}(age) \times HEI_0(P_n, age) \times NP(P_n, age) \right] \quad (5)$$

Here, $\Delta HEC$ denotes the absolute increase in cases of a given health effect relative to the baseline for each population group ($P_n$), defined by sex and country. $BAF$ represents the biological amplification factor associated with each health effect. The parameter $f_p$ is the local dose-reduction factor related to skin phototype (see Section 2.3.3). Age refers to the population cohort considered. $\Delta UV_{exp}$ is the cumulative percentage (or fractional) increase in lifetime exposure to biologically weighted UV radiation. $HEI_0$ is the baseline incidence rate of the health effect for each population and age group, and $NP$ represents the population size for each cohort and age group. The derivation of these variables is described in Section 2.3.4. Overall, this formulation estimates excess cases by applying cumulative, age-specific increases in biologically weighted ultraviolet exposure to baseline incidence rates across the population.

The standard operating assumption of this study is that individual behavioral patterns remain unchanged between the post-war and baseline scenarios. Governmental and societal responses to a nuclear war would introduce uncertainties that are difficult to quantify within the current modeling framework. Accordingly, human exposure behavior is assumed to remain constant over time, and potential adaptive or maladaptive responses are not explicitly represented. These include, but are not limited to, policy interventions, technological adaptations, weakened immune systems or deteriorating health conditions due to famine, increased public awareness of UV overexposure, limited access to healthcare, and changes in medical practices such as earlier detection and treatment of suspicious skin lesions.

*2.3.3 Skin Color as a Proxy for Dose Reduction*

Skin pigmentation provides varying levels of protection against erythema and skin cancer. To quantify this effect in people with Fitzpatrick skin types I-VI (Fitzpatrick, 1988), we used the UV sensitivity and MED values reported by D'Orazio et al. (2013) (Table 1).

Table 1. Skin pigmentation, the Fitzpatrick scale and UV risk adapted from D'Orazio et al. (2013).

| **Fitzpatrick phototype** | **Phenotype** | **Cutaneous response to UV** | **MED ($J/m^2$)** |
|---|---|---|---|
| I | Unexposed skin is bright white<br>Blue/green eyes typical<br>Freckling frequent<br>Northern European / British | Always burns<br>Peels<br>Never tans | 150-300 |
| II | Unexposed skin is white<br>Blue, hazel or brown eyes<br>Red, blonde or brown hair<br>European / Scandinavian | Burns easily<br>Peels<br>Tans minimally | 250-400 |
| III | Unexposed skin is fair<br>Brown eyes Dark hair<br>Southern or Central European | Burns moderately<br>Average tanning ability | 300-500 |
| IV | Unexposed skin is light brown<br>Dark eyes<br>Dark hair<br>Mediterranean, Asian or Latino | Burns minimally<br>Tans easily | 400-600 |
| V | Unexposed skin is brown<br>Dark eyes<br>Dark hair<br>East Indian, Native American, Latino or African | Rarely burns<br>Tans easily and substantially | 600-900 |
| VI | Unexposed skin is black<br>Dark eyes<br>Dark hair<br>African or Aboriginal ancestry | Almost never burns<br>Tans readily and profusely | 900-1500 |

However, the skin color classifications listed in Table 1 are not quantitatively linked to the incidence or mortality of UV-induced skin cancer, and they do not adequately represent the global distribution of skin phenotypes. In this study, we adopt the approach proposed by van Dijk et al. (2013) of using skin reflectance at 685 nm as a measure for skin color according to a linear model by Jablonski and Chaplin (2000):

$$ref = -0.2245 \times UV_{ery} + 74.4713 \quad (6)$$

Here, the $UV_{ery}$ is the 1990-2000 climatological mean UV dose distribution ($J/m^2$), weighted by CIE standard erythema spectrum using the TUV model. By applying this equation, we assumed that the local skin reflectance is adapted to the local UV climatology. Although this method does not explicitly represent the biological mechanisms through which pigmentation reduces skin cancer risk, nor fully capture the real-world global distribution of skin color, it currently provides the most practical and internally consistent framework for representing skin pigmentation as a proxy for UV dose attenuation at the global scale. The uncertainties associated with this assumption are further discussed in Section 4.2.

Then we estimate protector fact, $f_p$, using:

$$f_p = 10^{a\,(ref - ref_0)} \quad (7)$$

where $ref_0 = 66$ represents the reference skin reflectivity (66%) of "white" skin in Caucasian populations, for which the reference protection factor ($f_p$) is defined as 1. By fitting global skin cancer incidence data from Parkin et al. (2002), van Dijk et al. (2013) derived a parameter value of $a = -0.03935$. For dark skin with a reflectivity of 30% ($ref = 30$), the model yields a protection factor of $f_p = 26$. This estimate is broadly consistent with the approximately 33-fold higher MED reported for dark skin compared with light skin by Clydesdale et al. (2001).

*2.3.4 Baseline Incidence and Mortality*

The year 2000 is adopted as the baseline period in this study, including surface UV levels, global population distribution, and the demographic structure of each country by age and sex. Baseline melanoma and non-melanoma skin cancer mortality data are obtained from the Global Burden of Disease (GBD) dataset (Murray, 2022), which provides incidence and mortality estimates for multiple causes of disease, stratified by age, sex, and geographic location.

*2.3.5 Population Data.*

While the global population has continued to grow since the baseline year 2000, we assume a constant global population for all calculations. Population data disaggregated by sex, age, and country are required for the analysis. In this study, we use population data from the United Nations Department of Economic and Social Affairs, World Population Prospects 2024 (United Nations, 2024).

## 3 Results

### *3.1 Global distribution of UV-B*

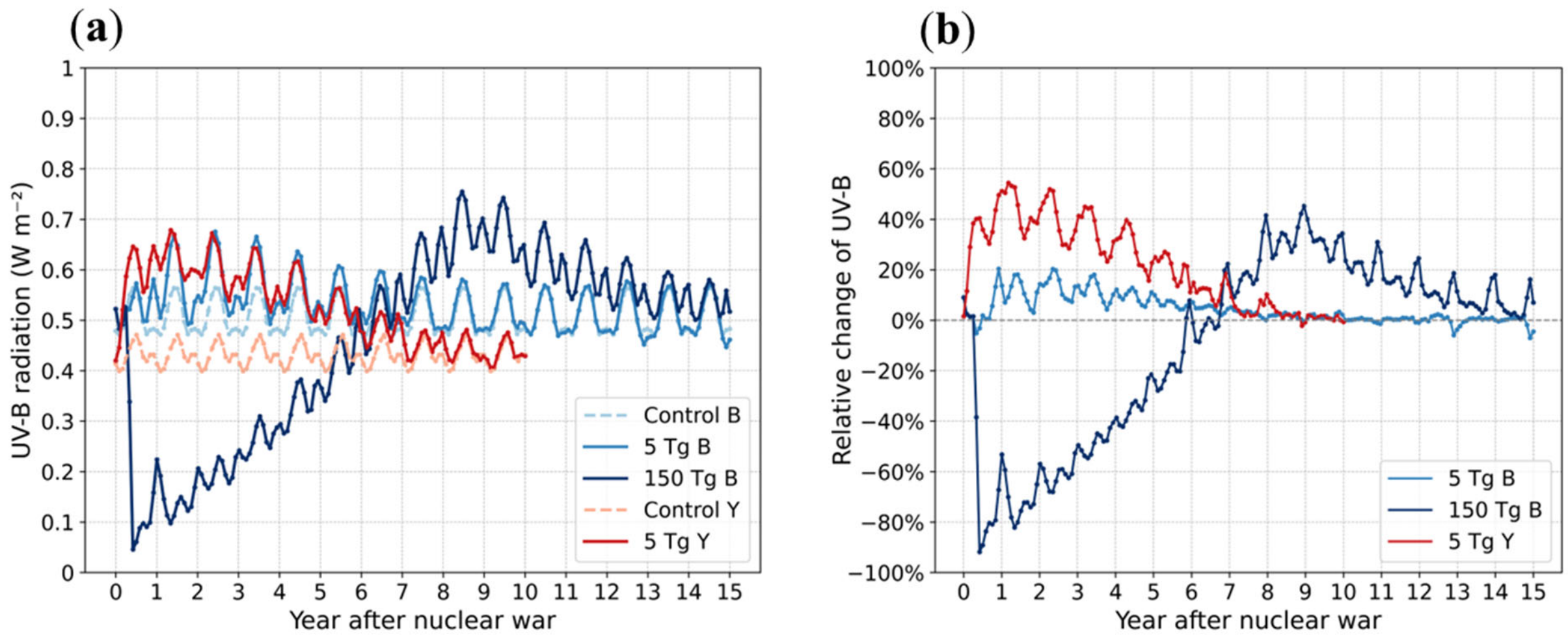

**Figure 2.** (a) Variation of UV-B under the 5 Tg, 150 Tg, and control scenarios from two studies, and (b) the corresponding relative changes with respect to the control in each study. B denotes Bardeen et al. (2021), and Y denotes Yook et al. (2025)

Simulations from Bardeen et al. (2021) initiate the nuclear war on May 15 in year 0, while Yook et al. (2025) started the war on January 11 in year 0, as is shown in Figure 2. Yook et al. (2025) also tested the emission during the Northern Hemisphere summer season, but found that the stratospheric cooling and ozone depletion were not sensitive to the timing of the emissions. In the control case, Bardeen et al. (2021) estimated a global mean UV-B ranging from 0.48 to 0.56 W $m^{-2}$ (Figure 2a). With updated chemistry and emissions, however, Yook et al. (2025) calculated a lower global mean UV-B, ranging from 0.40 to 0.48 W $m^{-2}$ (Figure 2a).

Following a regional nuclear war between India and Pakistan, both studies suggest a similar increase in UV-B after the explosions. UV-B peaks at about 0.68 W $m^{-2}$ (Figure 2a) and remains elevated through year 3, after which it gradually returns toward baseline levels. However, due to differences in the control simulations, the relative UV-B increase is larger in Yook et al. (2025) (up to 55%) than that reported by Bardeen et al. (2021) (approximately 20%). The immediate UV-B increase in the 5 Tg case is concentrated mainly in the mid- and high-latitude Northern Hemisphere (Figure S2). Small fluctuations near the end of both simulations likely reflect internal climate variability (Figure 2b).

In contrast, following a global war between United States and Russia, surface UV-B drops by roughly 90% within six months (Figure 2b), because soot injected into the stratosphere blocks incoming sunlight and sharply reduces UV transmission. As the soot heats the stratosphere, substantial ozone depletion occurs. A similar delayed UV response was found for the Chicxulub asteroid impact scenario (Ishida et al., 2007), in which large stratospheric aerosol loading initially offset the effect of ozone depletion on surface UV-B. As the aerosols declined over several years while ozone remained depleted, surface UV-B subsequently increased above pre-impact level, reaching approximately 1.5 times the unperturbed value under severe ozone depletion. This behavior is qualitatively similar to the transition from initial soot induced UV suppression to delayed UV enhancement in the 150 Tg nuclear war scenario. Model simulations suggest that the strong seasonal variability in Antarctic circulation and ozone allowed Southern Hemisphere surface UV-B to recover temporarily (Figure S1, S3), but this regional recovery did not produce a discernible increase in global mean UV radiation during the first several years after the global war. After the soot loading declines sufficiently, solar UV radiation reaching the surface will be enhanced. By years 8-9, surface UV-B becomes ~40% higher than in the control case.

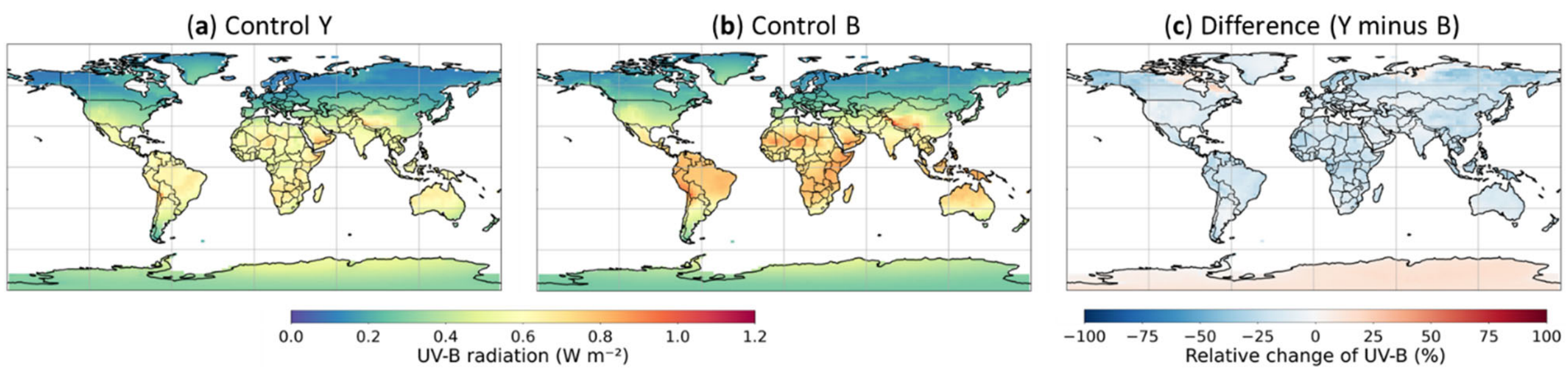

**Figure 3.** Annual mean distribution of global UV-B radiation under the control simulations from (a) Bardeen et al. (2021) and (b) Yook et al. (2025), and (c) relative UV-B changes between the two control cases.

Spatially, both studies show that the annual mean UV-B under control conditions is highest in the tropics and over the Tibetan Plateau, and lowest in the polar regions (Figures 3a, 3b). The UV-B field in Yook et al. (2025) is lower than that in Bardeen et al. (2021) over most of the globe, except for slightly higher values in parts of Antarctica (Figure 3c). In contrast, the relative increase in UV-B after a nuclear war is greatest at high latitudes and smallest in the tropics, particularly over central Africa and the Amazon Basin. We focus on the years of maximum UV-B enhancement: year 2 for the 5 Tg scenario (Figures 4a, 4b) and year 9 for the 150 Tg scenario (Figure 4c).

Under the 5 Tg scenario, Bardeen estimated that UV-B increases by ~60% in the Arctic and up to ~90% over Greenland (Figure 4b). Under the 150 Tg scenario, the Arctic experiences even larger increases, while UV-B over Antarctica rises by more than 100% due to severe ozone depletion (Figure 4c). However, even under a regional nuclear war scenario, Yook et al. (2025) found much stronger UV-B amplification (Figure 4a), with more than a twofold increase at latitudes above 45°N and across much of Antarctica. Regions with high population density, such as China, Europe, and the United States, would experience UV-B increases of around 50%. These widespread UV-B increases across both hemispheres demonstrate that altered UV exposure is a global consequence of nuclear war rather than a regionally confined effect.

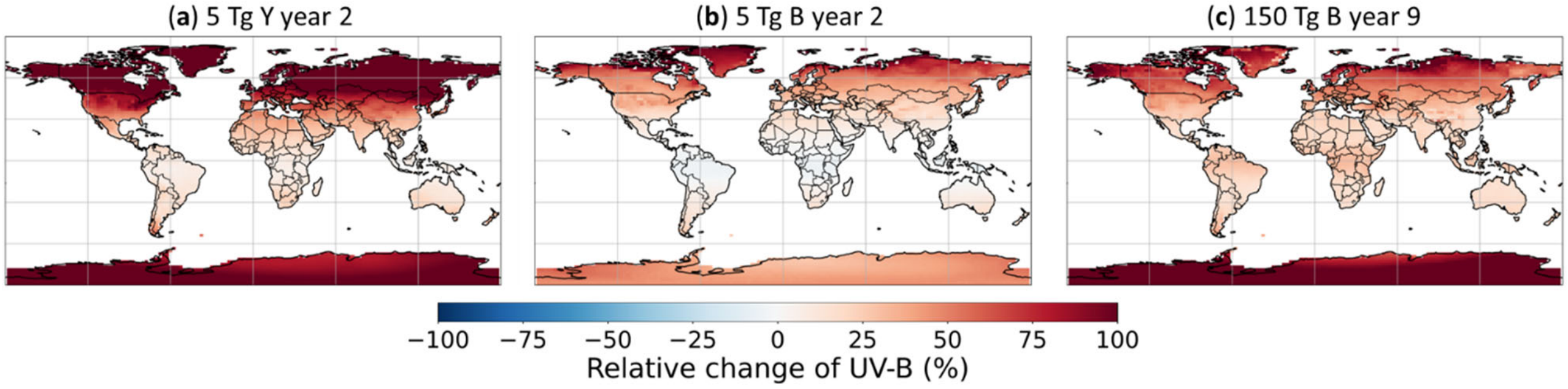

**Figure 4.** Relative UV-B changes under different nuclear war scenarios: the 5 Tg scenario in year 2 (a, Yook et al., 2025; b, Bardeen et al., 2021), and the 150 Tg scenario in year 9 (c, Bardeen et al., 2021).

***3.2 UV effects on sunburn exposure time***

Using the monthly global daytime mean UV-B radiation as an indicator of sunburn risk, we calculated the minimum exposure time required for individuals with different skin types to reach the average MED during years of maximum UV-B enhancement (Figure 5).

Globally, individuals with lighter skin types already experience very short safe exposure times under baseline conditions; therefore, post-war increases in UV-B do not substantially shorten these times further. However, the same exposure duration would result in a higher MED dose. Assuming 20 minutes of unprotected outdoor exposure under all the five scenarios, individuals with skin type I would receive the maximum value 6.1 times the MED in year 9 under the 150 Tg scenario, compared with the minimum value of 3.7 times the MED in year 2 of the Yook et al. control simulation. This increase corresponds to a 16% increase in the severity of inflammation caused by overexposure (Diffey, 2021).

During peak UV-B years, safe outdoor exposure times for skin types III and IV decrease to less than 15 minutes across most regions. Individuals with darker skin types, who typically tolerate longer exposure (Fitzpatrick, 1988), become increasingly vulnerable under extreme UV-B conditions. Under the 5 Tg scenario from Yook et al. and the 150 Tg scenario from Bardeen et al., the safe outdoor exposure time for individuals with skin type V and VI during summer decreases by approximately 10-20 minutes (Figures 5b, 5e).

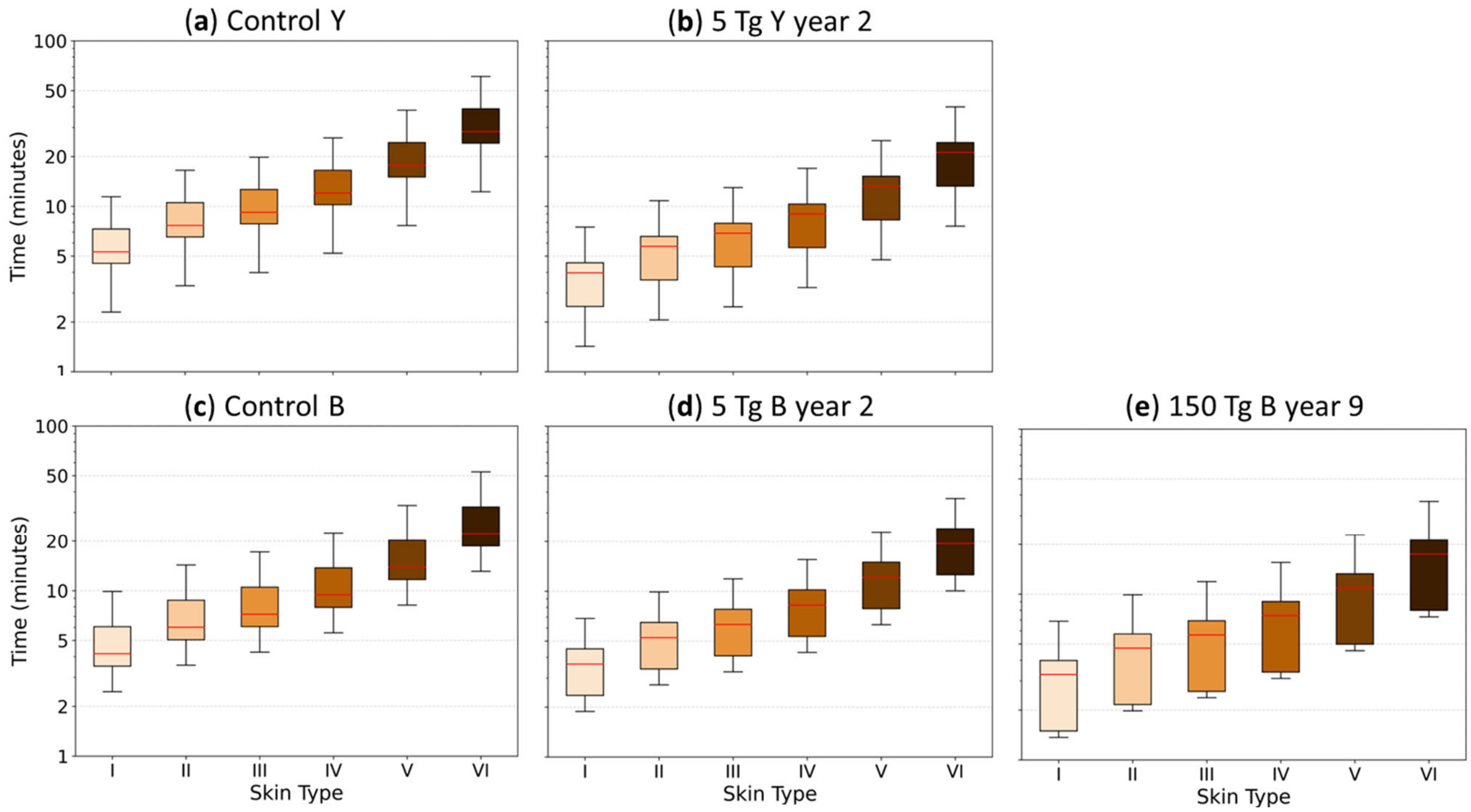

**Figure 5.** Box plots of global minimum exposure time for different skin types (see Table 1) under nuclear war scenarios. Panels a-b correspond to Yook et al. (2025) (control and 5 Tg in year 2), and panels c-e correspond to Bardeen et al. (2021) (control, 5 Tg in year 2, and 150 Tg in year 9).

To illustrate spatial patterns of exposure under a consistent biological sensitivity, Figure 6 is calculated assuming a globally uniform skin type II (Table 1), representative of relatively UV-sensitive populations (European or Scandinavian ancestry). To examine extreme conditions under each scenario, we used the highest surface UV-B radiation at each grid point during year 2 of the two 5 Tg scenarios (Figures 6a-d) and during year 9 of the 150 Tg scenario (Figures 6e, 6f).

However, because exposure is calculated using daytime mean UV-B, the estimated minimum exposure times represent daily averages and would vary over the course of the day, with the shortest safe exposure occurring near local noon. Safe outdoor exposure time decreases most strongly in polar regions, by more than 70%. Even though peak UV-B increases are smaller in densely populated regions such as Europe, North America, and China, individuals with lighter skin types are still not recommended to stay outdoors without protection for more than about 10 minutes. The extreme UV-B levels in high-latitude and high-altitude regions may therefore pose risks to both ecosystems and human societies.

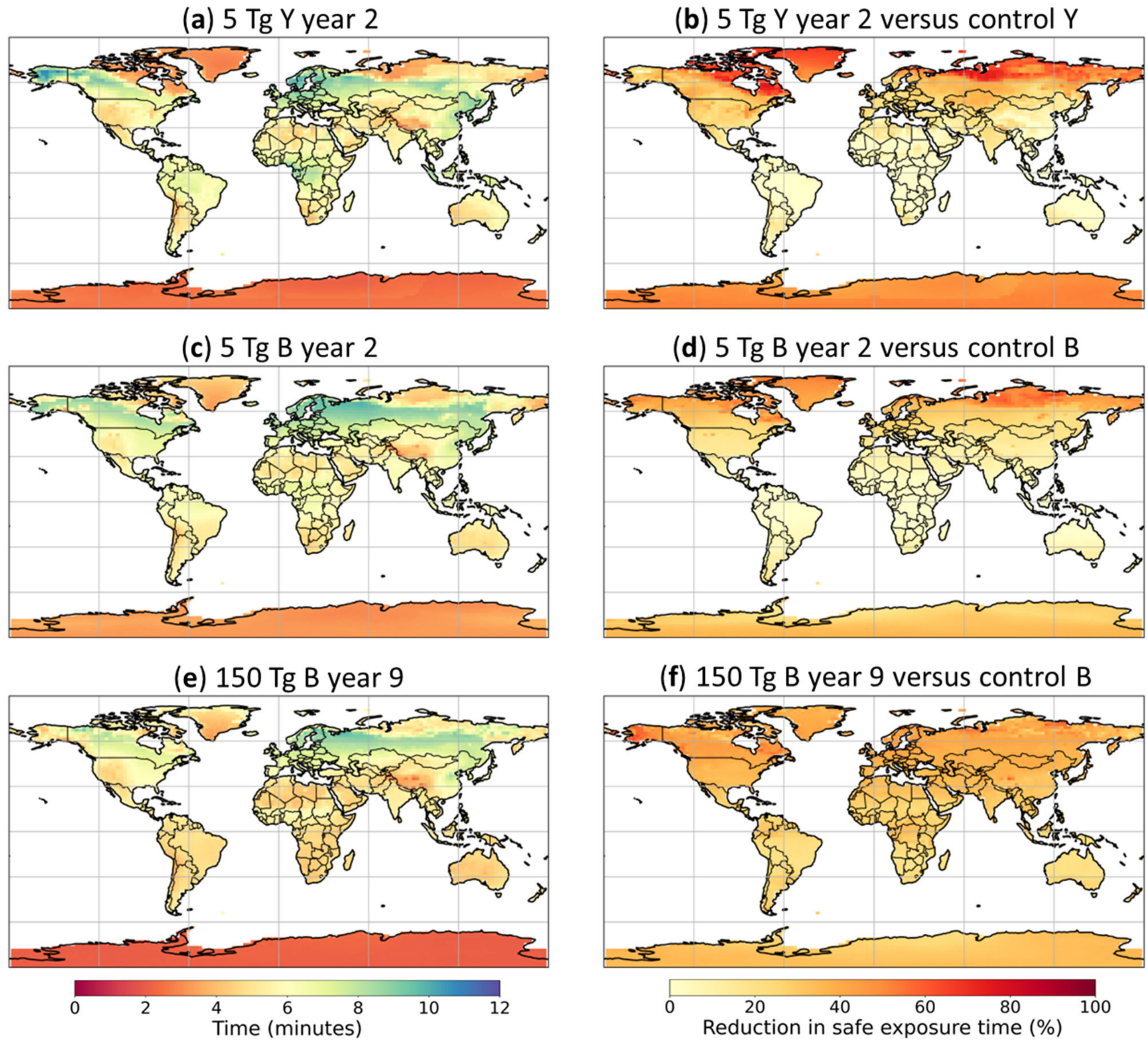

**Figure 6.** Safe exposure time for a globally uniform distribution of skin type II under nuclear war scenarios (left column) and the corresponding reduction relative to the control (right column). Panels a-b show results from Yook et al. (2025) for the 5 Tg scenario in year 2; panels c-d show the control and 5 Tg scenario (year 2) from Bardeen et al. (2021); panels e-f show the 150 Tg scenario in year 9 from Bardeen et al. (2021).

### *3.3 UV impacts on skin cancer mortality*

By the end of the nuclear war simulations, the global average cumulative excess mortality rates are estimated to be 0.06 for melanoma and 0.03 for NMSC per 100,000 people under the 5 Tg scenario, following 15 years of enhanced UV-B radiation according to Bardeen et al. (2021). Using the 10-year UV-B variations from Yook et al. (2025), we estimated higher global average cumulative excess mortality rates under the 5 Tg scenario, reaching 0.11 for melanoma and 0.05 for NMSC per 100,000 people. In contrast, under the 150 Tg scenario from Bardeen et al. (2021), the reduction in surface UV-B during the first six years outweighs the subsequent UV-B enhancement by the end of the 15-year simulation. Consequently, the global average cumulative mortality rates decrease by 0.09 for melanoma and 0.05 for NMSC per 100,000 people.

Spatially, changes in skin cancer mortality are highest in regions with large baseline mortality rates, large populations, and strong UV-B increases, particularly at higher latitudes, where populations are also more UV sensitive (Figure 7). Consequently, Russia, the United States, and China experience the largest variation in skin cancer mortality, coinciding with the countries that possess the largest nuclear arsenals.

Russia shows the greatest excess mortality due to its high latitude and large number of people with light skin color, with more than 600 and 1000 additional melanoma deaths (Figures 7a, 7b), and about 400 and 780 NMSC deaths (Figures 7d, 7e) under the two 5 Tg scenarios, respectively. Under the 150 Tg scenario, the initial suppression of UV-B results in approximately 740 fewer melanoma deaths and 460 fewer NMSC deaths by the end of the simulation (Figures 7c, 7f). The United States and China also experience significant changes in skin cancer mortality, driven by their large populations and high baseline mortality rates. In the United States, melanoma dominates the mortality variation, with approximately 400 and 950 additional deaths in the two 5 Tg scenarios, and about 940 fewer deaths in the 150 Tg scenario (Figures 7a-c). In China, NMSC contributes more strongly, with more than 300 and 570 additional deaths under the two 5 Tg scenarios and approximately 850 fewer deaths under the 150 Tg scenario (Figures 7d-f).

Besides these three nuclear-armed countries, Canada and Western Europe also show huge changes in skin cancer mortality. In contrast, African countries exhibit relatively small changes in mortality, largely due to smaller relative changes in UV-B and the generally darker skin pigmentation of their populations. Greenland, despite its high latitude and large UV-B increases, experiences very low mortality simply because of its small population.

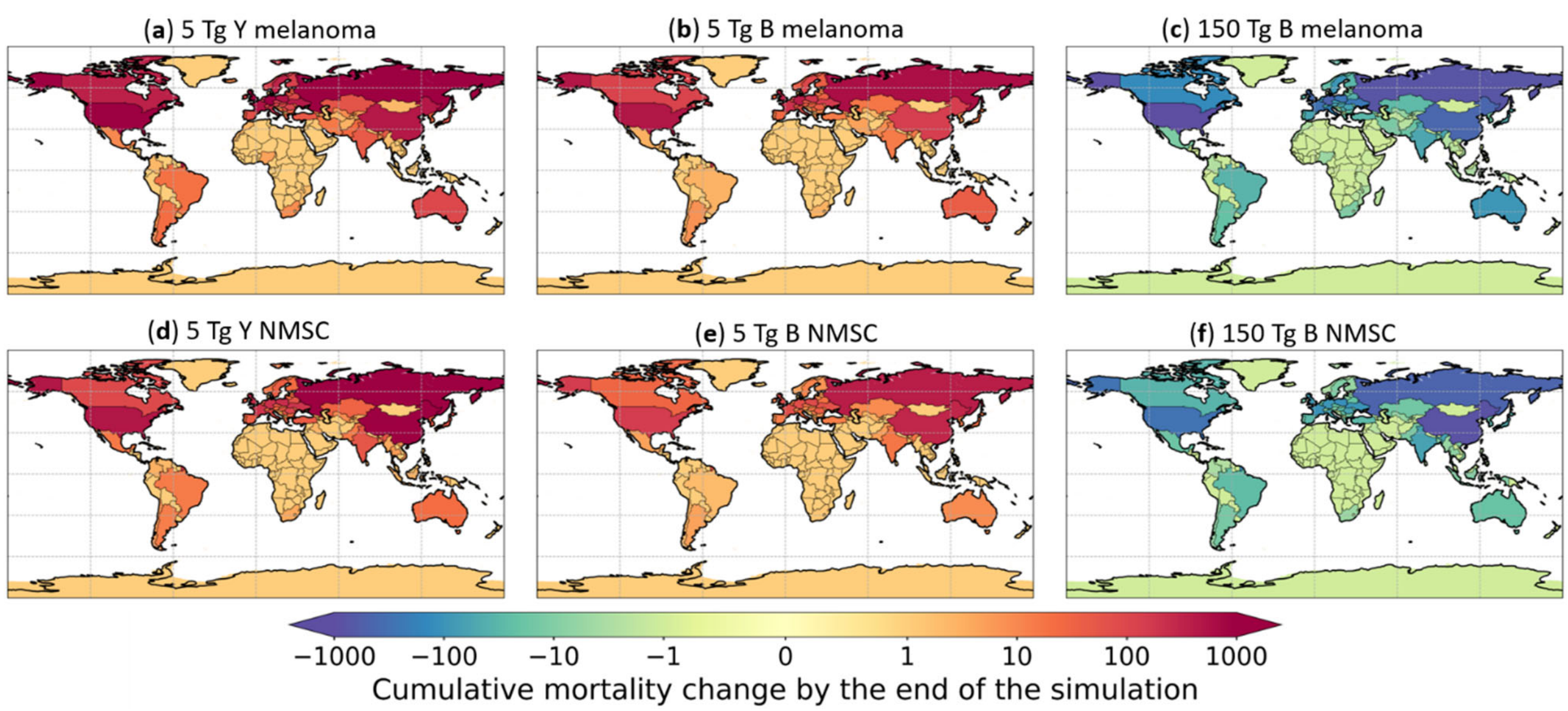

**Figure 7.** Changes of cumulative mortality number in each country by the end of different nuclear war scenarios. Panels (a) and (d) show results from Yook et al. (2025) for the 5 Tg scenario; panels (b) and (e) show the 5 Tg scenario from Bardeen et al. (2021); panels (c) and (f) show the 150 Tg scenario from Bardeen et al. (2021).

After 15 years and 10 years of cumulative UV-B exposure, we project approximately 5,300 and 9,800 excess UV-related skin cancer deaths globally under the 5 Tg scenarios from Bardeen et al. (2021) and Yook et al. (2025), respectively (Figure 8). The higher mortality estimates in the Yook et al. (2025) 5 Tg scenario are mainly caused by greater cumulative UV-B exposure and stronger UV-B increases across densely populated Northern Hemisphere regions. In contrast, the 150 Tg scenario from Bardeen et al. (2021) results in approximately 8,500 fewer skin cancer deaths globally by the end of the 15-year simulation, which is primarily driven by the suppression of surface UV-B during the first six years after the conflict.

As described in Section 2.3.2, the effects of even a few years of elevated UV exposure may persist over a lifetime (Slaper et al., 2001). This implies that even after surface UV radiation returns to baseline levels following the model simulations, individuals remain at an increased risk of developing skin cancer compared with a scenario without nuclear war. Accordingly, we assume that, after 10 or 15 years following each injection scenario, surface UV radiation returns to and remains at baseline levels for the remainder of the century. Based on this assumption, we further project the excess mortality attributable to the additional UV radiation induced by nuclear war over a 100-year period.

As shown in Figure 8, cumulative excess mortality continues to increase after year 15. By the end of the century, more than 21,200 additional deaths from skin cancer are projected under the 5 Tg scenarios from Bardeen et al. (2021). With updated chemistry, mortality in the 5 Tg scenario (Yook et al., 2025) could reach around 75,000 deaths. Under the 150 Tg scenario, the cumulative reduction in skin cancer mortality reaches a maximum of approximately 17,000 fewer deaths. As surface UV-B shifts from suppression to enhancement after year 7, the long-term effects of this increased exposure gradually offset the earlier reduction. As is shown in Figure 8, the cumulative mortality reduction begins to diminish around 70 years after the war. By the end of the century, the net reduction decreases to approximately 15,800 deaths. These projected totals are comparable to the baseline mortality rate in the year 2000 (1.19 per 100,000 people), corresponding to roughly 72,000 deaths annually, highlighting the persistent influence of nuclear war-induced UV changes on human health.

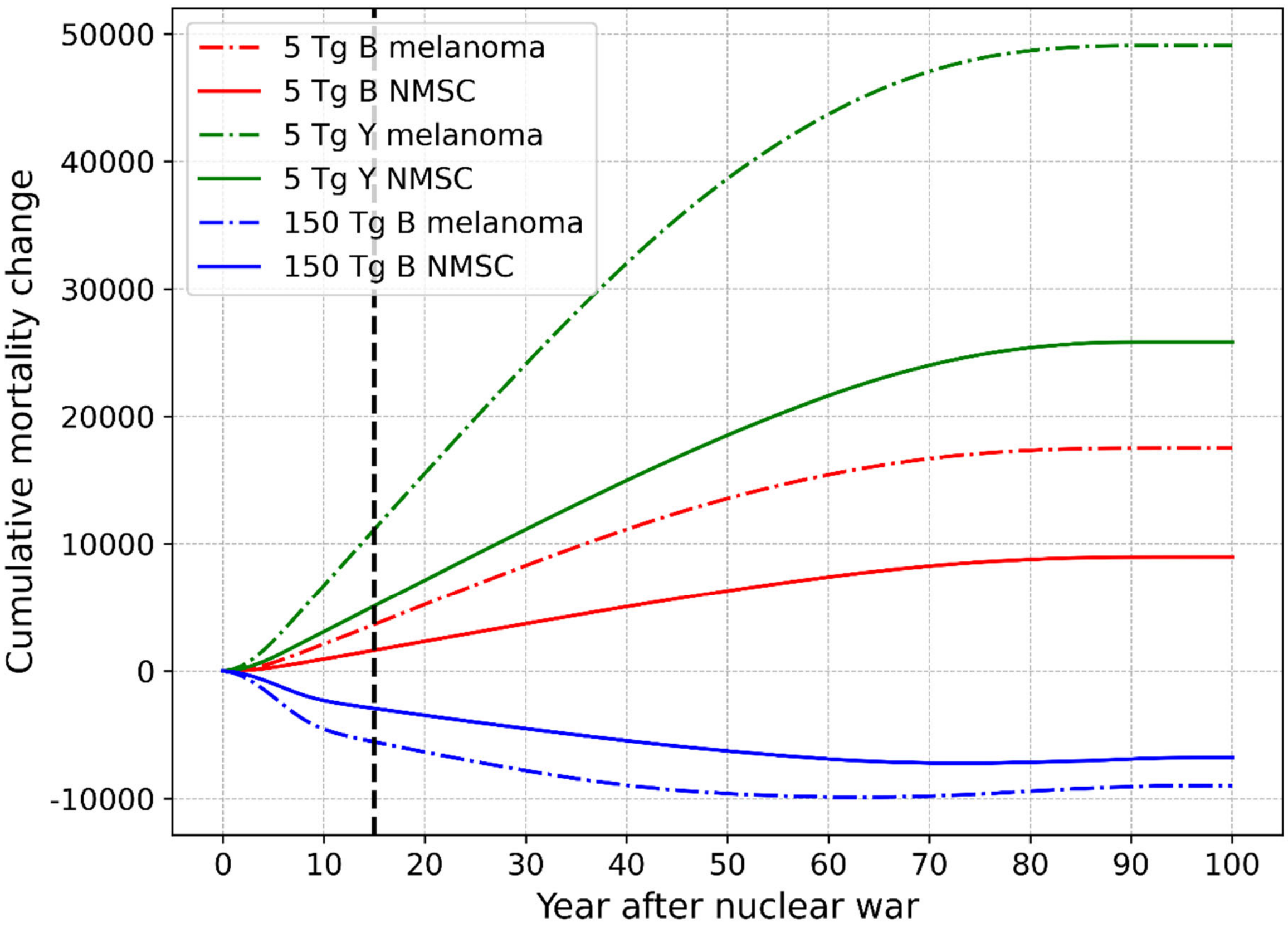

**Figure 8.** Global cumulative skin cancer mortality changes within one century after a nuclear war. The black dashed line indicates year 15, when surface UV levels return to normal in the Bardeen et al. (2021) simulations.

## 4 Discussion and Conclusions

### *4.1 Main findings and conclusions*

This study provides the first global assessment of health impacts from enhanced surface ultraviolet radiation following regional and global nuclear wars, using CESM-WACCM4 coupled with the TUV model and biologically-weighted UV exposure metrics. We show that the surface UV-B response to nuclear war is influenced by the competing influences of aerosol-induced attenuation and ozone depletion-driven enhancement, and that the direction and magnitude of net UV change depend critically on the scale of soot injection. Under the 150 Tg scenario, aerosol attenuation suppresses surface UV despite severe ozone depletion, whereas the 5 Tg regional scenario produces sustained net UV enhancement, with peak increases of 40-50%.

These UV changes would substantially alter safe outdoor exposure time and long-term skin cancer risk, with both the magnitude and direction of the health response depending on the scale of the conflict. Under the 5 Tg scenarios, we estimate thousands of additional skin cancer deaths within the first 15 years and 20,000-75,000 cumulative excess deaths over the subsequent century, driven by the cumulative nature of UV exposure and long-term carcinogenesis. The corresponding cumulative excess mortality rates range from 0.09 to 0.16 per 100,000 people within the first 15 years and from 0.43 to 1.23 per 100,000 over the century following the conflict, depending on the model configurations. These numbers are comparable to the baseline mortality rate in the year 2000 of roughly 72,000 deaths globally (1.19 per 100,000 people).

In contrast, under the 150 Tg scenario, strong soot attenuation of solar radiation results in 8,500 fewer skin cancer deaths within 15 years. The cumulative reduction subsequently reaches a maximum of approximately 17,000 deaths about 70 years after the explosion. As surface UV-B shifts from suppression to enhancement after year 7, the long-term effects of increased UV exposure gradually offset the earlier reduction, resulting in a net reduction of approximately 15,800 deaths by the end of the century. Correspondingly, the reduction in cumulative mortality increases from 0.14 per 100,000 people within the first 15 years to a maximum of 0.28 per 100,000 around year 70, before declining slightly to 0.26 per 100,000 by the end of the century.

An important finding of this study is that the surface UV response does not scale monotonically with the magnitude of a nuclear conflict. In the 5 Tg regional-war scenarios, the relatively smaller soot burden allows ozone-depletion-driven UV enhancement to reach the surface comparatively early. However, although the 150 Tg global-war scenario produces much more severe atmospheric disruption, dense stratospheric soot initially reduces global mean surface UV-B by approximately 90%. As the soot burden subsequently declines while ozone depletion persists, surface UV-B increases and reaches approximately 40% above the control by years 8-9 (Bardeen et al., 2021).

These contrasting responses imply different sequences of post-war UV hazards rather than a simple increase in UV impacts with war magnitude. Following a large-scale nuclear war, the early years would be dominated by reduced sunlight, extreme cooling, and severe food-system disruption, while enhanced UV would emerge later as atmospheric conditions begin to recover. Although our simulations suggest an overall reduction in skin cancer mortality under this scenario, the delayed UV enhancement could impose additional stress on human populations and ecosystems already weakened by the preceding climatic and food-system shocks. Under smaller regional-war scenarios, enhanced UV can become an important health and environmental stress much earlier. These results highlight the need to consider not only short-term human sunburn risks, but also the delayed long-term effects of altered UV radiation on ecosystems and human society.

### *4.2 Uncertainties and limitations*

#### *4.2.1 Uncertainties in sunburn quantification*

Sunburn, an inflammatory response of human skin, is a much faster biological process than skin cancer, typically occurring within several hours of excessive UV exposure (Neale et al., 2023). In addition to its role in the development of skin cancer, sunburn was included in our analysis to represent the short-term and immediate health effects associated with increased UV-B radiation following a nuclear war. However, quantifying the exact threshold at which sunburn occurs for an individual remains challenging. Some previous studies have attempted to apply thermal burn thresholds to UV-induced erythema, but this approach is inappropriate because UV-B radiation primarily affects deeper layers of the epidermis, whereas heat-induced burns damage more superficial layers (Altintas et al., 2009). Consequently, the term *"first-degree burn"* should not be used interchangeably for sunburn and superficial thermal injuries.

In this study, we instead use MED, a standard metric in dermatological research (D'Orazio et al., 2013). Although the values of MED vary across individuals and experimental protocol, they provide a biologically meaningful threshold for assessing relative changes in sunburn risk (Diffey, 2021). Reaching MED in natural sunlight typically produces mild erythema, a minimal sunburn rather than severe skin injury. Nevertheless, the widespread reduction in safe outdoor exposure time after a nuclear war would likely result in more frequent sunburn and sustained discomfort for large populations, representing an additional and often overlooked burden on daily life and outdoor activities.

*4.2.2 Skin pigmentation and behavior adaptation*

Skin pigmentation is an important source of heterogeneity in UV-related health impacts. Individuals with lighter skin pigmentation are more susceptible to UV-induced DNA damage and have higher risks of developing melanoma and NMSC than darker-skinned populations (D'Orazio et al., 2013). Assuming a globally Caucasian population results in approximately 3,100 and 7,000 additional UV-related skin cancer deaths under the 5 Tg scenarios of Bardeen et al. (2021) and Yook et al. (2025), respectively, and about 8,600 fewer deaths under the 150 Tg scenario of Bardeen et al. (2021). (Figure S5). Under this assumption, the United States and China populations would exhibit the largest increases in projected mortality. For the 5 Tg scenario, the estimated excess skin cancer deaths would increase by approximately 1,350 in China and 1,580 in the United States (Figure S5a, b). In contrast, the increase in Russia is relatively small because the previous simulations described in the manuscript already assume predominantly light-skinned populations there. These numbers are comparable to those obtained using the global distribution of skin pigmentation (Figure 7, 8), although the regional distribution of mortality would change substantially. This sensitivity analysis indicates that skin pigmentation is an important source of regional heterogeneity, while the overall mortality pattern reflects the combined effects of UV-B enhancement, population distribution, baseline skin cancer mortality, and population susceptibility.

In this study, we estimated the skin color dependent dose reduction using erythema-weighted UV doses from 1990 to 2000, essentially deriving a proxy for pigmentation based on local solar radiation (van Dijk et al., 2013). However, this approach does not fully represent real-world patterns of skin pigmentation, particularly in regions with large population migration. For example, populations in countries such as the United States have a wide range of ancestral UV environments (Monk, 2021). As a result, their skin pigmentation today does not match the UV climatology of their current location. A more accurate approach would involve mapping ethnic composition of each region and projecting it onto a high spatial global resolution, but such detailed demographic data are not available or practical for all countries.

Moreover, although our derived skin-color-dependent dose reduction aligns moderately with some established classifications, the spatial patterns might differ. For example, our method suggests that darker skin-color factors occur mainly in high-altitude populations, whereas in regions such as central and eastern Asia the estimated local phototypes fall into types I or II, which are typically associated with European or Scandinavian populations. According to Table 1, however, the actual skin phototypes in these regions are more likely to fall within types III to IV.

Lastly, our analysis assumes that individuals would not change their behavior or adopt additional protection—a common simplifying assumption in global modeling studies. However, cultural or adaptive behaviors complicate any static skin color representation. Individuals may tan, wear sunscreen, use protective clothing, or alter their exposure patterns. These adaptations would likely further change if people were informed that UV levels had increased after a nuclear war.

*4.2.3 Post war health conditions*

Extreme health stress after nuclear conflicts may increase individual susceptibility to skin cancer. For example, Collins et al. (2019) reviewed risk factors and treatment of skin cancer in immunosuppressed populations, including patients with HIV, autoimmune diseases, and solid-organ transplants, and those receiving long-term immunosuppressive therapy. Their results showed that immunosuppressed individuals would have a higher incidence of NMSC, particularly SCC. Gómez-Tomás et al. (2026) further examined the influence of different immunosuppressive regimens on skin cancer occurrence among solid-organ transplant recipients. During follow-up, 12.4% of participants developed skin cancer, and 450 separate skin cancer events were documented. Wang et al. (2020) also performed a meta-analysis of patients with psoriasis, and found that patients with psoriasis had approximately a 1.72-fold higher risk of NMSC than populations without psoriasis.

Changes in longevity may introduce an additional uncertainty. Increased life expectancy provides more time for cumulative UV exposure and for skin cancer to develop, potentially increasing lifetime skin cancer risk. However, population-level effects are difficult to quantify because poorer health may increase biological susceptibility while also shortening the period in which skin cancer can be diagnosed (Drejøe et al., 2026). They may die from other illnesses before a skin malignancy is detected. Existing datasets also lack the spatially and clinically resolved information needed to jointly represent multiple health conditions, UV exposure, and cancer outcomes. Therefore, although severe health stress may increase UV-related skin cancer risk among nuclear-war survivors, current evidence does not support a reliable quantitative adjustment, representing an important limitation and priority for future research.

*4.2.4 Post war demographic conditions*

Previous studies indicate that direct fatalities from nuclear conflict could reach tens to hundreds of millions, while climate-driven food shortages could place billions of people at risk. Toon et al. (2007) estimated more than 21 million direct deaths in a hypothetical India-Pakistan conflict involving 100 Hiroshima-sized weapons, whereas Xia et al. (2022) projected that 0.26-5.3 billion people could lack sufficient food two years after the war, depending on the soot-injection scenarios, food-system responses, and calculation assumptions. Under the severe population losses associated with the 150 Tg scenario in Xia et al. (2022), food insecurity would overwhelmingly dominate the overall health burden. After applying the corresponding reduced population fractions under the scenario where international food trade is turned off, 50% livestock grain feed is redirected to human consumption, and household food waste is made available for reuse, the estimated global reduction in skin-cancer mortality over 100 years decreases substantially by 94%, from about 17,000 to about 1,100 deaths. Compared with the catastrophic mortality due to blast, radiation, and famine, UV-related skin-cancer deaths are numerically much smaller.

However, under the 5 Tg scenario, adjusting the population for famine risk produces little change in either cumulative excess skin-cancer mortality or its spatial distribution (Figure S6 and Figure 7). This is consistent with the relatively limited effect of the 5 Tg food-production shock on the number of people that national food supplies could support in many regions. Russia, the United States, and China therefore remain among the countries with the largest numbers of excess skin-cancer deaths.

Although the cumulative deaths estimated here are small relative to the much larger and more uncertain mortality from explosions and famine, the aggregate excess mortality reported in this study—exceeding 75,000 deaths in some cases—is not trivial in absolute terms and represents deaths that would not otherwise occur. These results further emphasize that enhanced UV exposure may remain a meaningful long-term health burden following smaller-scale nuclear conflicts.

#### *4.2.5 Atmospheric-model and UV response uncertainty*

The systematic differences between the two model configurations applied in this research are not negligible compared with the simulated UV-B variability, which would contribute to the uncertainties in the simulation of skin cancer risk. The baseline difference is particularly important when interpreting the relative UV-B changes: although the absolute peak UV-B values under the 5 Tg scenarios are similar in the two studies, the lower control level of UV-B in Yook et al. (2025) results in a larger relative increase (~55%) than in Bardeen et al. (2021) (~20%).

Several factors may contribute to the differences in control UV-B climatology, including total column ozone, background atmospheric conditions such as cloud cover and temperature profiles, chemistry parameterizations, and model configuration. For example, global mean total column ozone in the control simulation ranged from 276 to 310 DU in Bardeen et al. (2021), compared with 276 to 300 DU in Yook et al. (2025). Differences in CESM versions, atmospheric chemistry configurations, and model dynamics may further affect ozone and, consequently, surface UV radiation. Given the numerous interacting chemical and dynamical processes involved, fully attributing the differences between the two simulations would require a dedicated model intercomparison with controlled sensitivity experiments, which is beyond the scope of this study. Here, we focus on the human health consequences of nuclear-war-induced UV changes, assuming that both modeling frameworks provide physically reasonable representations of post-war ozone and UV evolution. A systematic evaluation of the underlying chemical and dynamical mechanisms represents an important direction for future research.

### ***4.3 Broader implications of nuclear war induced UV changes***

#### *4.3.1 Evolution of estimates of post-nuclear war UV radiation*

Early assessments already recognized that nuclear-war-induced ozone depletion could produce enhanced surface UV after atmospheric smoke cleared. SCOPE 28 estimated a maximum ozone-column depletion of about 44% and found that a 40% ozone reduction at 45°N could increase UV-B by 38%, with much larger increases in DNA- and plant-weighted radiation (Hutchinson et al., 1985).

More recent models resolve this evolution more explicitly. Mills et al. (2014) projected substantial ozone loss and extreme UV Index values after a 5 Tg regional conflict, while Bardeen et al. (2021) showed that a 150 Tg global war could initially reduce surface UV-B by about 90% through soot attenuation, followed by an increase of about 40% above control by years 8-9 as the soot cleared. Together, these studies show that the basic sequence of initial UV suppression followed by delayed enhancement has remained consistent, while modern models provide much more detailed estimates of its magnitude, timing, and spatial distribution.

*4.3.2 Ecological and agricultural impacts*

Apart from the direct impacts on human health, enhanced UV radiation following a nuclear war could also have important implications for agriculture and livestock production. Although these effects are not explicitly quantified in this study, previous research indicates that elevated UV-B can influence animal health (e.g., Barnes et al., 2023).

Similar to humans, increased UV-B exposure may elevate the risk of skin and eye damage in livestock. Anderson and Badzioch (1991) reported a significant association between higher UV radiation levels and an increased incidence of ocular cancer in Hereford cattle in the United States and Canada. However, due to the limited availability of observational data and systematic studies, it remains difficult to quantify the global mortality or economic losses in livestock associated with enhanced UV-B radiation. Meanwhile, UV radiation may also have beneficial effects under controlled conditions. For example, Rana and Campbell (2021) showed that UV-A exposure can reduce fear and stress responses in livestock, while supplemental UV-B radiation may improve egg production and eggshell quality. Hodnik et al. (2025) further reported that UV-B exposure during automatic milking increased vitamin D levels in dairy cows and enhanced daily milk yield. However, such beneficial effects are largely dependent on controlled exposure conditions, and it remains unclear whether elevated UV levels under natural conditions would produce similar outcomes.

Crop productivity would be largely suppressed by the soot attenuation of incoming solar radiation during the first several years following a nuclear conflict. As the soot is gradually removed from the atmosphere and surface UV radiation increases, terrestrial vegetation could face an additional stress, particularly from UV-B, through molecular damage and changes in plant physiology and morphology (Liaqat et al., 2024).

Numerous field experiments have examined the effects of enhanced UV radiation on crops. Lizana et al. (2009), for example, found that increasing UV-B during the early vegetative-to-booting stage reduced aboveground biomass by 11-19% and grain yield by 12-20%. Using a modulated UV-B system maintained at 30% above ambient levels, Yin and Wang (2012) showed that UV-B exposure reduced maize height, with the strongest effect occurring during the elongation stage. Kataria and Baghel (2015) examined soybean performance under ambient UV conditions, and found that UV exposure constrained soybean productivity by reducing both carbon and nitrogen fixation. Mathur et al. (2024) further showed that UV-B negatively affected early-season morpho-physiological traits across 64 rice genotypes, including reductions in chlorophyll content and the nitrogen balance index.

Despite these evidences, few studies have examined how enhanced UV radiation may affect terrestrial ecosystems under the altered climatic conditions of nuclear winter. To date, the only directly relevant modeling study is Coupe et al. (2024), which used CESM2 to estimate UV inhibition of phytoplankton photosynthesis. A comprehensive assessment of nuclear-war consequences must therefore extend beyond the immediate effects of blast, famine, and climate disruption to include delayed impacts on both human and ecological systems. Enhanced UV-B exposure may increase long-term risks of sunburn and skin cancer, while also impairing terrestrial vegetation, crop productivity, livestock, and marine ecosystem. Most of these UV-related effects demand further study, and results in this research are smaller than the catastrophic near-term mortality associated with explosions and food-system collapse. However, they represent additional and potentially persistent burdens that would emerge as atmospheric soot declines and ozone depletion intensifies. Future studies should therefore integrate changes climate, population, and ecosystem responses within a bigger Earth-system and health-impact framework, then long-term consequences of nuclear conflict could be fully quantified.

**Acknowledgments**
This study was supported by the Future of Life Institute.

**Author Contributions**
S.X., L.X., and A.R. designed the study. C.B. and S.Y. conducted climate model simulations. S.X. analyzed the data with contributions from all the authors. S.X., L.X., and A.R. wrote the first draft and all authors contributed to editing and revising the manuscript.

**Conflict of Interest**
The authors declare no conflicts of interest relevant to this study.

**Data Availability Statement**
The climate model outputs used to assess health impacts are available from the studies of Bardeen et al. (2021) and Yook et al. (2025). Baseline melanoma and NMSC mortality data were obtained from the GBD dataset available at (https://vizhub.healthdata.org/gbd-results). Population data for the baseline year 2000 were obtained from the United Nations World Population Prospects 2024 (https://population.un.org/wpp). The TUV model outputs for the 1990 to 2000 climatological UV dose distribution are available from the https://www2.acom.ucar.edu.